\documentstyle[12pt]{article}
\normalbaselines
\begin{document}
\bibliographystyle{unsrt}
\vbox{\vspace{42mm}}

{\bf q--NONLINEARITY, DEFORMATIONS}\\
\indent{\bf AND PLANCK DISTRIBUTION}
{}~~\\
{}~~\\

\hskip 3cm {V. I. Man'ko$\star \dagger$, G. Marmo$\star$, and
F. Zaccaria$\star$}
\begin{center}
\begin{itemize}\item[~$\star$]Dipartimento di Scienze Fisiche,\\
Universit\'a di Napoli "Federico II"\\
Istituto Nazionale di Fisica Nucleare, Sezione di Napoli\\
Mostra d'Oltremare, Pad.19 - 80125 Napoli, Italy\\
\item[~$\dagger$]Lebedev Physics Institute\\
53 Leninsky Prospekt, 117924 Moscow, Russia\end{itemize}
\end{center}
{}~~\\
{\bf INTRODUCTION}
{}~~\\

In this contribution we will review a new approach to some nonlinear
dynamical systems which is related to the q--deformation (or other types
of deformations) of linear classical and quantum systems considered in
\cite{sol1}. The main idea of this approach is to replace constants like
frequency or mass, {\it etc.,} which are parameters of the linear systems
with constants of the motion of the system. This procedure produces
from the initial linear system a nonlinear one and as it was demonstrated
in \cite{sol1}, \cite{sol2} the q--oscillator of \cite{bie}, \cite{mc} may
be considered as physical system with this specific nonlinearity which was
called q--nonlinearity. At the example of q--oscillator the constant
parameter which was replaced by the constant of the motion dependent on
the amplitude of the vibration was the frequency.

But this approach may be used for many dynamical systems. So, the constant
masses in Klein--Gordon and Dirac equations or the constant signal velocity
in the wave equation may be replaced by the dynamical variables but
these variables are chosen to be the constant of the motion of the
dynamical system under consideration. Conceptually we address the problem
(which is not new) if the physical constant parameters like light velocity,
Planck constant or gravitational constant are constant parameters
in reality or they may change, for example, in the process of
evolution? The same question may be put about such characteristics of
elementary particle as electric charge or mass, say, of electron or fine
structure constant which play the role of constant parameters in the
dynamical equations of the theory. The example of the q--nonlinearity
of \cite{sol1} showed that a linear equation or system of equations may be
considered as a limit of nonlinear system characterized by a nonlinearity
parameter which is not important for small intensities but starts to play
important role for the high field intensities or for large energy
densities.

Taking into account such phenomenon naturally yields the
deformation of the linear equations of the motion which are transformed
due to the deformation into the nonlinear ones but of specific type. The
corresponding nonlinear system of equations is simply a "reparametrized"
initial linear system of equations in which constant parameters are
replaced by the constants of the motion of the nonlinear system. This
specific procedure of nonlinearization of the initial linear system of
equations selects a broad enough class of integrable dynamical systems
which are simple but could describe the mentioned above physical phenomena.
Also such dynamical systems are appropriate to treat the q--deformations
and other aspects of quantum qroups \cite{drin}, \cite{jimbo},
\cite{wor} and quantum
qeometry \cite{manin} as influence of different types of
nonlinearities in the corresponding physical processes.

Of course, the formulated approach may be extended in the sense
that one could deform not only initial linear system to make it the
nonlinear one using method of replacing constant parameters by constants
of the motion. In principle, one could deform a simple initial
nonlinear system containing some constant parameters and to transform
it into another nonlinear system by replacing the constants with the
integrals of the motion of the nonlinear system. So, it is possible to
extend the class of integrable nonlinear systems starting from simple
integrable nonlinear systems and reparametrizing them, i. e. replacing
the constant parameters by the constants of the motion. Such approach
may be usefull, for instance, in q--deforming general relativity or
Yang--Mills type gauge theories. Up to our knowledge such classes of
both linear and nonlinear deformed integrable dynamical systems have
not been studied till now. It should be noted that the method of
obtaining the integrable nonlinear systems starting from the linear
ones is close in spirit to the generalized reduction procedure
of considering the integrable dynamical systems \cite{manmar},
\cite{lan}.

The goal of our work is to present a general scheme for the described
procedure of deforming the linear systems with finite number of degrees
of freedom (next Section). In subsequent sections we also will give
some examples of the deformation including infinite number of degrees
of freedom (wave equation). Then the
physical consequencies of the q-- nonlinearity as blue shift effect
\cite{sol2}, deformation of Planck distribution formula
\cite{sol2}, \cite{sol3}, and change of a charge form--factor
\cite{marman} will be discussed. In the last two sections deformed
Klein--Gordon equation and Maxwell
equation in the vacuum will be considered.
{}~~\\
{}~~\\
{}~~\\
{\bf SPECIAL NONLINEAR SYSTEMS OBTAINED FROM LINEAR ONES}
{}~~\\

We consider a carrier space $~{\cal R}^{n}$. With any matrix $~A$ we
associate the vector field $~X_{A}=~x_{i}A_{j}^{i}
\partial /\partial x_{j}$. This association is a Lie algebra
isomorphism. By using the associative product of matrices, we can
associate powers with any matrix $~A~~~~~A~\cdot ~A\cdots ~A=~A^{k}$
for any integer $~k$. The number of independent ones is given by the
degree of the minimal polynomial of $~A$. If we associate vector fields with
any one of these powers we get
\begin{equation}
X_{k}=~(A^{k})_{n}^{m}x_{m}\frac {\partial }{\partial x_{n}}
\end{equation}
and
\begin{equation}
[X_{k},~X_{j}]=0,
\end{equation}
therefore we get mutually commuting symmetries for our initial dynamical
system $~X_{A}$. We notice that for a generic matrix $~A$ there will be
$~n$--independent matrices and they are a basis of all infinitesimal
symmetries for $~X_{A}$ with coefficients constants of the motion for
$~X_{A}$.

For nonsingular matrices $~A^{k}$, i. e. $~\det (\hbox {\bf 1}
+A^{k})\neq 0$, we can define invertible transformations
\begin{equation}
\Phi _{k}=\frac {\hbox {\bf 1}-A_{k}}{\hbox {\bf 1}+A_{k}}
\end{equation}
which are symmetries for $~X_{A}$, i. e.
\begin{equation}
\Phi _{k}^{-1}A\Phi _{k}=A,
\end{equation}
or we can consider $~e^{sA^{k}}=\varphi _{k}$, which are also symmetries
for $~X_{A}$.

When $~A$ is not generic, there are other infinitesimal symmetries
which are not obtained from powers of $~A$ and moreover the algebra of
symmetries need not be Abelian. Again we can use either the Cayley map
or the exponentiation to get finite symmetries out of the infinitesimal
ones.

Now we can make the dynamics and the infinitesimal symmetries nonlinear
by replacing the entries $~A_{j}^{i}\in {\cal R}$ with constants of the
motion for $~X_{A}$. With this procedure, infinitesimal symmetries are no
more linear and they do not close on a finite dimensional Lie algebra.

If $~{\cal F}_{A}\in {\cal F}({\cal R}^{n})$ is the subring of constants
of the motion for $~X_{A}$, and $~X_{1},~X_{2},~\ldots ,~X_{S}$ is a basis
of linear infinitesimal symmetries for $~X_{A}$, we get nonlinear
infinitesimal symmetries  $~X_{f}=~f^{1}X_{1}+f^{2}X_{2}+\cdots +f^{S}X_{S}$
for any choice of $~f^{1},~f^{2},~\ldots ~,~f^{S}\in {\cal F}_{A}$. If
$~X_{A}$ is "reparametrized", i. e. replaced with $~f~X_{A},
{}~~~f\in {\cal F}_{A}$, previous symmetries are broken except for those which
preserve $~f$, i. e. $~\angle _{X_{j}}f=~0$.

To make contact with the quantum situation we shall consider previous
procedure in the framework of Hamiltonian dynamics.
If we consider the carrier space $~{\cal R}^{2n}$, with the symplectic
matrix
\begin{equation}
J=\left (\matrix{0&\hbox {\bf 1}\cr
-\hbox {\bf 1}&0\cr}\right )
\end{equation}
the dynamics represented by $~X_{A}$ will be Hamiltonian if
$~^{t}A~J+~J~A=~0$. It follows that $~J~A$ is a symmetric matrix and we
can define a function $~f_{A}(x)=^{t}xJAx$. This function turns out
to be the Hamiltonian function giving rise to $~X_{A}$.
In constructing infinitesimal symmetries for $~X_{A}$ by taking powers
of $~A$, not all powers will give rise to matrices in the symplectic
Lie algebra, we have to restrict to odd powers, i. e.
$~A^{1},~A^{3},~\ldots ,~A^{2k+1},~\ldots ,~A^{2n+1}$, indeed
\begin{equation}
^{t}(A^{2k+1})J+JA^{2k+1}=0
\end{equation}
and we find functions
\begin{equation}
f_{k}(x)=^{t}x(JA^{2k+1})x.
\end{equation}
{}From $~[A^{2k+1},~A^{2j+1}]=0$ and the Lie algebra isomorphism
$$~\{f_{B},~f_{C}\}=f_{[B,C]}$$
when $~f_{B}(x)=^{t}xBx$, we get $~\{f_{k},~f_{j}\}=0$, therefore
odd powers of $~A$ give rise to constants of the motion which are in
involution. For a generic matrix $~A,~~~X_{A}$ turns out to be a
completely integrable system in the Liouville sense.

The even powers of $~A$ give rise to symmetries which are not canonical.
We can turn them into nonlinear ones by using the procedure we have already
illustrated previously.
To deal with nonlinear transformations it is better to work in a geometrical
framework. Therefore on $~{\cal R}^{2n}$ we have the dynamics
$~X_{A}=~x_{i}A_{j}^{i}\partial /\partial x_{j}$, the simplectic structure
$~\omega =\sum _{i}dx_{i}\wedge dx_{i+n}$ and canonical symmetries
$~X_{1},~X_{2},~\ldots ,~X_{n}$ associated with odd powers of $~A$ and
functions $~f_{1},~f_{2},~\ldots ,~f_{n}$, while $~Y_{1},~Y_{2},~\ldots ,
{}~Y_{n}$ associated with even powers of $~A$ and are noncanonical. By
exponentiating the noncanonical infinitesimal symmetries we get invertible
transformations from $~{\cal R}^{2n}$ into $~{\cal R}^{2n}$ which takes us
from one Hamiltonian description for $~X_{A}$ to a different Hamiltonian
description.

As a final comment from linear algebra we recall that any matrix $~A$ can
be decomposed into the sum
\begin{equation}
A=~S+N
\end{equation}
where $~S$ is semisimple and $~N$ is nilpotent, moreover $~[S,N]=0$.
{}From here it follows that the flow associated with $~A$ can be written as
\begin{equation}
e^{tA}=~e^{tS}~e^{tN}.
\end{equation}
This formula shows that if we want (non trivial) bounded motions associated
with one
Hamiltonian vector field $~A$, we have to impose $~N=0$ and $~S$ should have
purely imaginary (non zero) eigenvalues. Therefore a generic linear
Hamiltonian system with bounded motions must be a collection of
noninteracting harmonic oscillators. To be more specific we turn to
consider a two--dimensional harmonic oscillator.
{}~~\\
{}~~\\
{}~~\\
{\bf THE HARMONIC OSCILLATOR}
{}~~\\

On $~{\cal R}^{4}$ with coordinates $~(~\tilde x,~\tilde v)$ we consider
the equations of motion
\begin{equation}
\frac {d\tilde x_{k}}{dt}=\tilde v_{k},~~~~~\frac {\tilde v_{k}}{dt}=
-\omega ^{2}_{k}\tilde x_{k},~~~~~k=~1,~2.
\end{equation}
We introduce new coordinates
\begin{equation}
q_{k}=\tilde x_{k},~~~~~p_{k}=\frac {\tilde v_{k}}{\omega _{k}},~~~~~
k=~1,~2
\end{equation}
and get
\begin{equation}
\frac {dq_{k}}{dt}=\omega _{k}p_{k},~~~~~\frac {dp_{k}}{dt}
=-\omega _{k}q_{k}
\end{equation}
associated with the matrix
\begin{equation}
A=\left (\matrix{0&\omega _{1}&0&0\cr
-\omega _{1}&0&0&0\cr
0&0&0&\omega _{2}\cr
0&0&-\omega _{2}&0}\right ).
\end{equation}
We find the commuting symmetries
\begin{equation}
A^{0}=\left (\matrix{1&0&0&0\cr
0&1&0&0\cr
0&0&1&0\cr
0&0&0&1}\right ),
\end{equation}
\begin{equation}
A^{2}=\left (\matrix{-\omega _{1}^{2}&0&0&0\cr
0&-\omega _{1}^{2}&0&0&0\cr
0&0&-\omega _{2}^{2}&0\cr
0&0&0&-\omega _{2}^{2}}\right ),
\end{equation}
and
\begin{equation}
A^{3}=\left (\matrix{0&-\omega _{1}^{3}&0&0\cr
\omega _{1}^{3}&0&0&0\cr
0&0&0&-\omega _{2}^{3}\cr
0&0&\omega _{2}^{3}&0}\right ).
\end{equation}
A more convenient basis of symmetries gives rise to the vector fields
\begin{eqnarray}
X_{1}=q_{1}\frac {\partial }{\partial p_{1}}
-p_{1}\frac {\partial }{\partial q_{1}}&,&
X_{2}=q_{2}\frac {\partial }{\partial p_{2}}
-p_{2}\frac {\partial }{\partial q_{2}},\nonumber\\
Y_{1}=q_{1}\frac {\partial }{\partial p_{1}}
+p_{1}\frac {\partial }{\partial q_{1}}&,&
Y_{2}=q_{2}\frac {\partial }{\partial p_{2}}
+p_{2}\frac {\partial }{\partial q_{2}}.
\end{eqnarray}
Associated functions are
$$h_{1}=p_{1}^{2}+q_{1}^{2},~~~~~h_{1}=p_{2}^{2}+q_{2}^{2}.$$
By using $~Y_{1}$ and $~Y_{2}$ made into nonlinear transformations,
we can consider
\begin{eqnarray}
x_{k}&=&f_{k}(h_{k})p_{k},\nonumber\\
y_{k}&=&f_{k}(h_{k})q_{k}.
\end{eqnarray}
This nonlinear transformation is a symmetry for the dynamics but is
noncanonical, it means
\begin{equation}
\sum _{k}dx_{k}\wedge dy_{k}=\sum _{k}d(f_{k}p_{k})\wedge d(f_{k}q_{k})
\end{equation}
which is different from $~\sum _{k}dp_{k}\wedge dq_{k}$ giving rise to
standard Hamiltonian description of our harmonic oscillator with
Hamiltonian
$$H=\sum _{k}\frac {\omega _{k}}{2}(p_{k}^{2}+q_{k}^{2}).$$

The picture we are presented with is the following:\\
$~X_{A}$ has Hamiltonian description given by $~(\omega ,~H)$ or
$~(\tilde \omega ,~\tilde H)$, i. e.
\begin{equation}
i_{X_{A}}\omega =-dH,~~~~~
i_{X_{A}}\tilde \omega =-d\tilde H.
\end{equation}
What happens if we associate a vector field with $~\tilde H$
by using $~\omega $? i. e. we consider
\begin{equation}
i_{X}\omega =-d\tilde H,
\end{equation}
it turns out that $~X$ is a reparametrization of $~X_{A}$ by a constant
of the motion.

At this point we can consider the coordinate system where $~\omega $ has
the standard form (Heisenberg coordinates) and $~\tilde H$ is a deformation
of the quadratic expression or we can use a coordinate system where
$~\tilde H$ has the standard quadratic expression but $~\omega $ is
"deformed" (via the nonlinear noncanonical transformation). As $~\omega $
gives rise to Poisson brackets ("commutation relations") this second choice
can be interpreted via a deformation of the commutation relations and is
the starting point for "q--deformed oscillators."

Introducing complex coordinates in $\bf R^{4}$ we consider complex coordinates
\begin{equation}
\alpha_{k} = x_{k} + iy_{k},~~~~~~~~~\alpha^{*}_{k} = x_{k} - iy_{k}.
\end{equation}
The complex structure in $\bf R^{4}$ is given by the matrix
$$
J = \left ( \begin{array}{clcr}
0 & 1 & 0 & 0\\
-1 & 0 & 0 & 0\\
0 & 0 & 0 & 1\\
0 & 0 & -1 & 0\end{array}\right )
$$
satisfying $J^{2}=-1$. This matrix defines a complex structure commuting
with the dynamical evolution
\begin{equation}
\Gamma = i (\alpha_{k} \frac {\partial}{\partial \alpha_{k}} - \alpha^{*}_{k}
\frac {\partial}{\partial \alpha^{*}_{k}}).
\end{equation}
In the $(p,q)$ coordinates we have new complex coordinates
\begin{equation}
\xi_{k} = p_{k} + i q_{k},~~~~~~~~\xi^{}_{k} = p_{k} - i q_{k},
\end{equation}
and, setting
\begin{equation}
n_{k} = \alpha_{k} \alpha^{*}_{k},
\end{equation}
\begin{equation}
\xi_{k} = f_{k}(n_{k}) \alpha_{k},~~~~~~~\xi^{*}_{k} =
f_{k}(n_{k}) \alpha^{*}_{k}.
\end{equation}
Thus transformation is not "analytic" (we notice that analyticity depends
on the complex structure and here we have two alternative complex structures
compatible with the dynamics).
In these complex coordinates, if we take the new point, steming from qantum
mechanics, which takes the Hamiltonian as a primitive concept for the
dynamics, we are naturally led to consider the following two Hamiltonians
\begin{equation}
H_{1} = \frac {1}{2} \sum_{k} n_{k}
\end{equation}
in the $\alpha$-coordinates and
\begin{equation}
H_{2} = \frac {1}{2} \sum_{k} \xi_{k} \xi^{*}_{k}
\end{equation}
in the $\xi$-coordinates.
To compare, we express them in the same variables to find
\begin{equation}
H_{1} = \frac {1}{2} \sum_{k} n_{k}
\end{equation}
and
\begin{equation}
H_{2} = \frac {1}{2} \sum_{k} f_{k}^{2}(n_{k}) n_{k}.
\end{equation}
We now use the same bracket for them, say
\begin{equation}
\{\alpha_{k}, \alpha^{*}_{k}\} = i.
\end{equation}
We obtain two different dynamical systems, where the one associated
with $H_{2}$ is even not necessarily isotropic. The evolution goes
from periodic orbit to orbit whose closure is a 2-dimensional torus,
i.e. the associated systems are completely different. Of course, there
is no contradiction. Indeed to have the same dynamics we should use
different Poisson Brackets, namely, the one we get transforming Eq. (31).
It is now clear that under quantization these complex coordinates go into
creation and annihilation operators. Therefore for the corresponding
commutators we can repeat what we have said for the Poisson Bracket.

As a final remark, we notice that the notion of coherent states
is a kinematical notion. The coherent states are discussed in quantum
mechanics from the point of view of the states the closest to classical
one. According to this physical property the coherent states of the
electromagnetic field harmonic oscillator were introduced in \cite{gla}.
The coherent states (classical Gaussian packets) of a charge moving in
magnetic field were introduced in \cite{malm68}, \cite{feld}.
Coherent states of spin were introduced in \cite{rad}.

We start with a Poisson bracket compatible with
a complex structure, we consider a canonical chart to be anyone such
that
\begin{equation}
\{\chi_{k}, \chi^{*}_{j}\} = i \delta_{kj},
\end{equation}
then we associate creation and annihilation operators with these variables
and define the associated coherent states.
It is clear therefore that the ingredients needed for coherent states
are:

1) a symplectic structure

2) a compatible complex structure\\
and together they imply the existence of an Hermitian structure.
When there are various structures  compatible with a given dynamical
evolution, we find various coherent states (i.e.associated with various
symplectic structures) preserved by this evolution. The discussed complex
structure has been used to construct Jordan--Schwinger map for some
functional groups in \cite{patr}.
{}~~\\
{}~~\\
{}~~\\
{\bf CLASSICAL q--OSCILLATOR AS "REPARAMETRIZATION"}
{}~~\\

In previous section we described the general scheme of deformations. Now we
give in detail the example of q--oscillator. Also, the q--oscillator
mathematical properties and its modifications have been used to associate
with some physical phenomena \cite{usd}--\cite{chai}.
The q--oscillators are interesting from the point of view of mathematical
structure \cite{fair}--\cite{spir}.
Let us start from the considering the one--dimensional linear harmonic
oscillator with frequency $~\omega $ and corresponding nonlinear q--oscillator
following the approach of Ref. \cite{sol1}. The linear second order equation
for the coordinate of this oscillator $~x$ is
\begin{equation}
\ddot x+\omega ^{2}x=0.
\end{equation}
In the form of the first order equations for the position and momentum this
harmonic oscillator is described by the system
\begin{eqnarray}
\dot p&=&-\omega ^2x,\nonumber\\
\dot x&=&p.
\end{eqnarray}
The variables $~x$ and $~p$ are real ones. If we introduce complex variables
\begin{eqnarray}
\alpha &=&\frac {1}{\sqrt 2}(\sqrt {\omega }x
+\frac {i}{\sqrt \omega }p),\nonumber\\
\alpha ^{*}&=&\frac {1}{\sqrt 2}(\sqrt {\omega }x
-\frac {i}{\sqrt \omega }p)
\end{eqnarray}
the harmonic oscillator is described by the first order differential
equations system
\begin{eqnarray}
\dot \alpha &=&-i\omega \alpha ,\nonumber\\
\dot \alpha ^*&=&i\omega \alpha ^*.
\end{eqnarray}
Both systems (34) and (35) may be rewritten in matrix form of Ref.
\cite{sol1}
\begin{equation}
\dot{\underline{X}} = A \underline{X}.
\end{equation}
The real two--vector $\underline{X} \in \hbox {\bf R}^2$ and the matrix
$~A$ for the system (34) are
\begin{equation}
\underline {X}=\left (\matrix{p\cr
x\cr}\right ),~~~~~A=\left (\matrix{0&-\omega ^{2}\cr
1&0\cr}\right ).
\end{equation}
The complex vector $\underline{X}$ and matrix $~A$ for the system (35)
are
\begin{equation}
\underline {X}=\left (\matrix{\alpha \cr
\alpha ^{*}\cr}\right ),~~~~~A=-i\omega \left (\matrix{1&0\cr
0&-1\cr}\right ).
\end{equation}
According to our general procedure let us replace the
constant matrix elements $~A_{i}^{j}$ of matrix $~A$ by functions
of the constants of the motion. The system obtained is nonlinear and
it may be integrated by exponentiation as easily as the initial linear
system. We illustrate this procedure again on the harmonic oscillator
equations of motion (39) since in this case the matrix $~A$ is diagonal
one. Thus the constant frequency $~\omega $ should be replaced by the
constant of the motion $~\widetilde \omega (\alpha ,\alpha ^{*})$ and
we have new system for one and the same variable $~\alpha $
\begin{eqnarray}
\dot \alpha &
=&-i\widetilde \omega (\alpha ,\alpha ^{*})\alpha ,\nonumber\\
\dot \alpha ^{*}&=&i\widetilde \omega (\alpha ,\alpha ^{*})\alpha ^{*}.
\end{eqnarray}
If one takes the constant of the motion $~\widetilde \omega (\alpha ,
{}~\alpha ^{*})$ to be integral of motion independent explicitly on time
we have the condition for this function
\begin{equation}
\frac {d}{dt}\widetilde \omega (\alpha ,\alpha ^{*})
=\frac {\partial \widetilde \omega }{\partial \alpha }\dot \alpha +
\frac {\partial \widetilde \omega }{\partial \alpha ^{*}}\dot \alpha ^{*}
=0
\end{equation}
which after using the equation of motion (40) may be integrated and  the
solution for this equation is
\begin{equation}
\widetilde \omega (\alpha ,\alpha ^{*})
=\Omega (\alpha \alpha ^{*}),
\end{equation}
i. e. any function $\Omega $ of the modulus of the amplitude is the
frequency which is the constant of the motion. We will take this function
in the form
\begin{equation}
\Omega =\omega f_{q}(\alpha \alpha ^{*}).
\end{equation}
The function $~f_{q}(\alpha \alpha ^{*})$ will specify the q--deformation
of the harmonic oscillator and we will take it in the form
\begin{equation}
f_{q}(z)=\frac {\lambda }{\sinh \lambda }\cosh \lambda z,
{}~~~~~q=e^{\lambda },
\end{equation}
where $~\lambda $ is the parameter of q--nonlinearity of the vibrations.
In case $~\lambda \rightarrow 0$ the function $~f_{q}(z)\rightarrow 1$.
The relation to classical q--oscillator (and its quantum counter part)
consists of the existence the nonlinear noncanonical transform in the
phase space of the oscillator which in the variables $~\alpha $ looks
\cite{sol1} as
\begin{equation}
\alpha _{q}=\sqrt {\frac {\sinh \lambda \alpha \alpha ^{*}}
{\alpha \alpha ^{*} \,
\sinh \lambda }} \alpha ,~~~~~~~\alpha ^{*}_{q} =
\sqrt {\frac {\sinh \, \lambda \alpha \alpha ^{*}}
{\alpha \alpha ^{*} \, \sinh \lambda }} \alpha ^{*}.
\end{equation}
The Poisson Brackets of the variables $~\alpha _{q}$ reproduce the
structure of the annihilation and creation operators and the
commutation relations of the q--oscillator \cite{bie} (see, below).
The dynamics of the physical variables $~\alpha $ due to the
Hamiltonian
\begin{equation}
H=~\omega \alpha _{q}^{*}\alpha _{q},
\end{equation}
which is taken in form to be the same as the Hamiltonian of the
harmonic oscillator
\begin{equation}
H=~\omega \alpha ^{*}\alpha ,
\end{equation}
is the nonlinear vibration corresponding to Eq. (40). The linear dynamics
of the variable $~\alpha $ corresponding to the Hamiltonian (47) is
given by the solution to the Eq. (36). It is nessesary to point out that
in the suggested approach the physical meaning of the variables
$~\alpha $ as related to physical position $~x$ and physical momentum $~p$
in the sense of the measurement procedure by the formula (35) is preserved
for nonlinearized dynamics. It means that the dynamics of the coordinate
$~x$ and momentum $~p$ of the nonlinear q--oscillator is described by the
system of equations
\begin{eqnarray}
\dot x&=&f_{q}(\alpha \alpha ^{*})p,\nonumber\\
\dot p&=&-\omega ^{2}f_{q}(\alpha \alpha ^{*})x
\end{eqnarray}
where
\begin{equation}
\alpha \alpha ^{*}=\frac {1}{2}(\omega x^{2}+\frac {1}{\omega }p^{2}).
\end{equation}
For the deformed equations of motion of q--oscillator as the formula (48)
shows the momentum is the function of the velocity and position. This
function may be obtained as the solution to the functional equation
\begin{equation}
p(x,~\dot x)=\frac {\sinh \lambda }{\lambda }\frac {\dot x}
{\cosh [\frac {\lambda }{2}\{\omega x^{2}
+\frac {1}{\omega }p^{2}(x,~\dot x)\}]}
\end{equation}
considered as the implicit formula for the giving momentum as the function
of the position $~x$ and velocity $~\dot x$. The solution to q--oscillator
equation of motion
\begin{equation}
\dot \alpha =-i\omega \alpha \frac {\lambda }{\sinh \lambda }\cosh \lambda
\alpha \alpha ^{*}
\end{equation}
is
\begin{equation}
\alpha (t)=\alpha _{0}\exp [-i\omega t\frac {\lambda }{\sinh \lambda }
\cosh \lambda \alpha _{0}\alpha _{0}^{*}]
\end{equation}
where
$$\alpha _{0}=\alpha (t=0)$$
is the initial complex amplitude of the nonlinear q--oscillator. The
solution to the equation of motion for the coordinate $~x$ of the nonlinear
q--oscillator
\begin{equation}
\ddot x+\omega ^{2}\frac {\lambda ^{2}}{\sinh^{2}\lambda }
\cosh ^{2}\{\frac {\lambda }{2\omega }[x^{2}\omega ^{2}+p^{2}(x,~\dot x)]\}=0
\end{equation}
where the function $~p(x,~\dot x)$ is given explicitly by the relation (50),
may be written in the form
\begin{eqnarray}
x(t)&=&\frac {x_{0}}{2}\{\exp[\frac {i\lambda \omega t}{\sinh \lambda }
\cosh \{\frac {\lambda }{2\omega }[x_{0}^{2}\omega ^{2}
+p^{2}(x_{0},~\dot x_{0})]\}]\nonumber\\
&+&\exp[-\frac {i\lambda \omega t}{\sinh \lambda }
\cosh \{\frac {\lambda }{2\omega }[x_{0}^{2}\omega ^{2}
+p^{2}(x_{0},~\dot x_{0})]\}]\}\nonumber\\
&+&\frac {\dot x_{0}\sinh \lambda }{2i\lambda \omega }\cosh ^{-1}
\{\frac {\lambda }{2\omega }[x_{0}^{2}\omega ^{2}
+p^{2}(x_{0},~\dot x_{0})]\}\nonumber\\
&\times &\{\exp [\frac {i\lambda \omega t}{\sinh \lambda }
\cosh \{\frac {\lambda }{2\omega }[x_{0}^{2}\omega ^{2}
+p^{2}(x_{0},~\dot x_{0})]\}]\nonumber\\
&-&\exp [-\frac {i\lambda \omega t}{\sinh \lambda }
\cosh \{\frac {\lambda }{2\omega }[x_{0}^{2}\omega ^{2}
+p^{2}(x_{0},~\dot x_{0})]\}]\}.
\end{eqnarray}
Here $~x_{0}=~x(t=0)$ and $~\dot x_{0}=~\dot x(t=0)$ are the initial
position and velocity of the nonlinear q--oscillator. In the limit
$~\lambda \rightarrow 0$ we have the usual solution for the linear
harmonic oscillator.

One can find for small nonlinearity $~\lambda \ll 1$ the approximate
expression for the momentum solving the equation (50) by iteration method.
We have
\begin{equation}
p=\dot x[1+\frac {\lambda ^{2}}{6}-\frac {\lambda ^{2}}{8}(\omega x^{2}
+\frac {\dot x^{2}}{\omega })^{2}].
\end{equation}
This formula may be interpreated as the negative shift of the mass of the
oscillator by the factor depending quadratically on the energy of the
oscillations.
{}~~\\
{}~~\\
{}~~\\
{\bf TWO--DIMENSIONAL CLASSICAL q--OSCILLATOR}
{}~~\\

The dynamics of generically deformed two--dimensional oscillator is
described in previous sections. In this Section we will give the formula
for specific q--deformation of this oscillator \cite{sol1}. If one
considers two degrees of freedom, the amplitudes of the first harmonic
oscillator $~\alpha _{+}$ and of the second oscillator $~\alpha _{-}$ may
be q--deformed by two different methods. One is to have frequency of the
first oscillator to be dependent only on the energy of this oscillator
$~n_{+}=|\alpha _{+}|^{2}$ and the frequency of the second oscillator to be
dependent only on its energy $~n_{-}=|\alpha _{-}|^{2}$. Another method of
deformation is to make the frequencies of both oscillators to be dependent
on the full energy of vibrations
$$n=n_{+}+n_{-}.$$
We now will describe the q--deformed variables $~\alpha _{{q}_{\pm }}$ in
the case of q--deformation when the frequency of vibrations depends on full
energy. Then we have the following non--zero Poisson Brackets
\begin{equation}
\{\alpha_{q\pm}, \alpha^*_{q\mp}\} = i \alpha_{\pm}\alpha^*_{\mp} \frac
{(\lambda n)\cosh n\lambda - \sinh n\lambda }{n^{2} \sinh \lambda },
\end{equation}
and
\begin{equation}
\{\alpha_{q\pm},\alpha^*_{q\pm}\}
= \frac {i}{n \sinh \lambda}[ (1 - \frac
{n_{\pm}}{n})\sinh n\lambda + \lambda n_{\pm} \cosh n\lambda ].
\end{equation}
Here
\begin{equation}
\alpha _{{q}_{\pm }}=\alpha _{\pm }F(n)
\end{equation}
where
\begin{equation}
F(n)=\sqrt {\frac {\sinh \lambda n}{n \, \sinh \lambda }}.
\end{equation}
The introduced deformation of two harmonic oscillators implies the
interaction of these oscillators which exists due to q--nonlinearity.
Thus we have not only selfinteraction of the different modes but the
mutual influence of the motion of one oscillator onto the other.
{}~~\\
{}~~\\
{}~~\\
{\bf DEFORMED WAVE EQUATION}
{}~~\\

Now we consider the case of the system with infinite number of degrees
of freedom. An example of such a system is the wave equation with
the constant parameter which is wave velocity. This constant parameter
we take to be unity. Thus we start from the wave equation of the form
\begin{equation}
(\frac{\partial ^2}{\partial t^2}-
\frac{\partial ^2}{\partial x^2})\varphi (x,t)=0.
\end{equation}
In order to clarify the deformation procedure we represent this equation
as system of equations for decoupled oscillators. To do this we rewrite
this equation in momentum representation
\begin{equation}
\ddot \varphi (k,t)+k^2\varphi (k,t)=0
\end{equation}
where the complex Fourier amplitude
\begin{equation}
\varphi (k,t)=\frac{1}{2\pi }\int \varphi (x,t)\exp (-ikx)~dx,
\end{equation}
plays the role of new coordinate. Since $~\varphi (x,~t)
=~\varphi ^{*}(x,~t)$ we have
$~\varphi (k,~t)=~\varphi ^{*}(-k,~t)$.
Eq. (61) describes a two--dimensional oscillator with equal frequencies
for both modes labelled by $~k$ and $~-k$. Writing the Eq. (61) in the form
\begin{eqnarray}
\dot \varphi (k,~t)&=&\pi (k,~t),\nonumber\\
\dot \pi (k,~t)&=&-k^{2}\varphi (k,~t)
\end{eqnarray}
we have the equations of the form (38) in which frequency
$~\omega ^{2}=k^{2},~p\rightarrow \pi (k,~t)$ and
$~x\rightarrow \varphi (k,~t)$. Due to this we can deform this linear
system taking the integral of the motion
\begin{equation}
\mu=\int dk\{\frac {1}{2|k|}[k^{2}|\varphi |^{2}(k,t)
+|F_{k}|^{2}]\}
\end{equation}
in which the function $~F_{k}$ playing the role of complex momentum of
$~k$--field mode is solution to the infinite system of
equations
\begin{equation}
\dot \varphi (k,t)=F_{k}f_{q}\{\int dk'\frac {1}{2|k'|}
[k^{'2}|\varphi |^{2}(k',t)+|F_{k'}|^{2}]\}.
\end{equation}
The function $~f_{q}$ is given by the Eq. (44). Then the parameter
$~\mu $ plays the role of the initial number of vibrations corresponding
to given Cauchy initial conditions.

Thus we made the deformation by method used in the previous Section for
two--dimensional oscillator introducing the interaction of modes through
the parameter $~\mu$. The deformed wave equation may be written as
\begin{equation}
\ddot \varphi (x,t)
=-\int k^{2}f_{q}^{2}(\mu )\varphi (k,t)e^{ikx}dk.
\end{equation}
This equation may be rewritten in the form for which only scalar field
in time--space coordinate $~\varphi (x,t)$ is present. For this we
replace in the integrand the Fourier components $~\varphi (k,t)$  by its
expression in terms of $~\varphi (x,t)$ as well as in the integral of
motion $~\mu $ which, in principle, may depend on the momentum vector
$~k$ through the dependence on the Fourier components of the initial
field $~\varphi (x,0)$ and $~\dot \varphi (x,0)$. Thus we have the
deformed wave equation
\begin{equation}
\ddot \varphi (x,t)=-\frac {1}{2\pi }
\int \int k^{2}\exp [ik(x-x')]\varphi (x',t)
f_{q}^{2}\{\mu [\varphi (x,0),\dot \varphi (x,0)]\}~dk~dx',
\end{equation}
We have pointed out here that the integral of motion $~\mu $ is the
functional of the field $~\varphi (x,t)$ of the special form. Being the
integral of motion it depends only on initial values of field and field
velocities.

Since the parameter $~\mu $ is chosen as the common integral of motion
for all field oscillators it does not depend on wave vector $~k$,
and the function $~f_{q}^{2}(\mu )$ can be considered as a commom factor.
Then the wave equation may be rewritten in the form
\begin{equation}
\ddot \varphi (x,t)=f_{q}^{2}(\mu )\frac {\partial ^{2}}{\partial  x^{2}}
\varphi (x,t).
\end{equation}
We have the differential--functional equation which looks like usual
wave equation with the wave velocity $~f_{q}(\mu )$ which is constant of
the motion. Thus the procedure of deformation yields us the nonlinear
equation for which the velocity of wave propagation depends on the initial
configuration of the field and its time derivative. This equation
may be appropriate to describe a field behaviour in a media with strong
nonlinear response of its properties to the presence of the quanta of
the field.

Now we will obtain the solutions to nonlinear deformed wave
equation (68). The parameter $~\mu$ behaves as constant
for any choice of the initial conditions $~\varphi (x,t=0),~~
\dot \varphi (x,t=0)$. Due to this the solution to the nonlinear
mode--vibration equation (61) is
\begin{eqnarray}
\varphi (k,t)&=&\frac{1}{2}\varphi (k,0)\{\exp [i|k|f_{q}(\mu )]t+
\exp [-i|k|f_{q}(\mu )]t\}\nonumber\\
&+&\frac{1}{2i}\dot \varphi (k,0)\{\exp [i|k|f_{q}(\mu )]t
-\exp [-i|k|f_{q}(\mu )]t\}\frac {1}{i|k|f_{q}(\mu )},
\end{eqnarray}
and
\begin{eqnarray}
\varphi (k,0)&=&\frac{1}{2\pi }\int
\varphi (x,0)\exp (-ikx)~dx,\nonumber\\
\dot \varphi (k,0)&=&\frac{1}{2\pi }\int
\dot \varphi (x,0)\exp (-ikx)~dx.
\end{eqnarray}
It is multimode generalization of the solution (54) of the one--mode
nonlinear oscillator. Thus given initial condition
$~\varphi (x,0),~~\dot \varphi (x,0)$ implies that $~\varphi (k,0)$
and $~\dot \varphi (k,0)$, and $~\mu$ are given, too. In terms of
these values we have the $~k$--th mode solution $~\varphi (k,t)$ and
the solution to nonlinear q--deformed wave equation (68) are
given by Eqs. (62), (69). It is easy to prove that the q--deformed
wave equation (68) has the soliton--like solutions
\begin{equation}
\varphi _{\pm }(x,~t)=~\Phi (x\pm f_{q}(\mu )t)
\end{equation}
where $~\Phi $ is an arbitrary function. In fact, discussed
q--deformation implies the existence of nonlinear interaction
among the modes. The generalization to the case of three space
coordinates may be performed following the same reparametrization
procedure. The q-deformed Klein--Gordon equation is considered by this
method in \cite{zac}. From the point of view of deformed Poincare
symmetry group the relativistic equations are discussed in \cite{luk},
\cite{pil}.
{}~~\\
{}~~\\
{}~~\\
{\bf  QUANTUM q--OSCILLATOR}
{}~~\\

To make more clear the relation of the suggested approach to classical
equations of motion with standard quantum q--oscillator formalism of
\cite{bie} we review the description of the oscillator given in \cite{sol1}.
Let us introduce the usual creation and annihilation oscillator operators
$a$ and $a^{\dag}$ obeying bosonic commutation relations
\begin{equation}
[a,a^{\dag}]=1.
\end{equation}
Below we assume the classical dynamical variables to which $~a$
and $~a^{\dag}$ correspond to oscillate with a frequency $~\omega =~1$.
It is known that the operators $~a,~a^{\dag},~1$ form the Lie algebra of
Heisenberg--Weyl group. So, the linear harmonic oscillator  may be connected
with the generators of pure Heisenberg--Weyl Lie group. In view of the
commutation relation (72) the usual scheme for generating the states of the
harmonic  oscillator is based on the properties of the Hermitean number
operator  $~\hat n=~a^{\dag}~a$
\begin{equation}
[a,\hat{n}]=a,~~~[a^{\dag},\hat{n}]=-a^{\dag}.
\end{equation}
Thus
constructing the vacuum state $~|0 \rangle$ obeying the equation
\begin{equation}
a|0 \rangle =0,
\end{equation}
and the excited states
\begin{equation}
|n \rangle =\frac{a^{\dag~n}} {\sqrt{n!}}|0 \rangle
\end{equation}
which are eigenstates of the number operator $~\hat{n}$
\begin{equation}
\hat{n}|n \rangle =n|n \rangle, n \in Z^{+}
\end{equation}
the matrix representation of the operators $~a$ and $~a^{\dag}$ in the
basis (75) have the known expressions
\begin{eqnarray}
a&=&\left(\begin{array}{crcl}
0 & \sqrt{1} & 0 & \ldots \\
0 & 0 & \sqrt{2} & 0 \\
0 & 0 & 0 & \sqrt{3} \\
\ldots & \ldots & \ldots & \dots
\end{array}\right),\nonumber\\
a^{\dag}&=&\left(\begin{array}{crcl}
0 & 0 & 0 & \ldots \\
\sqrt{1} & 0 & 0 & \ldots \\
0 &\sqrt{2} & 0 & \ldots \\
\ldots & \ldots & \ldots & \ldots
\end{array}
\right)
\end{eqnarray}
while the number operator $~\hat{n}$ is described by the matrix
\begin{equation}
\hat{n}=\left(\begin{array}{crcl}
0 & 0 & 0 & \ldots \\
0 & 1 & 0 & \ldots \\
0 & 0 & 2 & \ldots \\
. & . & . & \ldots \\
\end{array}\right).
\end{equation}
The Hamiltonian for such a system is defined as
\begin{equation}
H = \frac{a^{\dagger}a + aa^{\dagger}}{2}.
\end{equation}
The q--oscillators may be introduced by generalizing
 the matrices (77) and (78)
with the help of the q--integer numbers $~n_{q}$,
\begin{equation}
n_{q}=\frac{\sinh~n\lambda }{\sinh~\lambda },~~~q=e^{\lambda }.
\end{equation}
Here $~\lambda $ and $~q$ are dimensionless $~c$-numbers,
which appear at this purely mathematical level.
When $~\lambda =0,~q=1$ and the $~q$--integer $~n_{q}$ coincides with
$~n$. Then, replacing the integers in (77) and (78)
by $~q$--integers we obtain  matrices which define  the annihilation
and creation operators of the quantum q-oscillator,
\begin{eqnarray}
a_{q}&=&\left(\begin{array}{crcl}
0 &  \sqrt{1_{q}} & 0 & ...\\
0 & 0 & \sqrt{2_{q}} &...\\
0 & 0 & 0& \sqrt{3_{q}}\\
...&...&...&...
\end{array}\right),\nonumber\\
a^{\dag}_{q}&=&\left(\begin{array}{crcl}
0&0&0&...\\
\sqrt{1_{q}}&0&0&...\\
0&\sqrt{2_{q}}&0&... \\
...&...&...&...
\end{array}\right),\nonumber\\
\hat{n}_{q}&=&\left(\begin{array}{crcl}
0&0&0&...\\
0&1_{q}&0&...\\
0&0&2_{q}&...\\
...&...&...&...
\end{array}\right),
\end{eqnarray}
since the action of $\hat n_q$ on eigenstates $|n>$ is given by
\begin{equation}
\hat n_q |n> = \frac {\sinh n\lambda }{\sinh \lambda }|n>.\nonumber\\
\end{equation}
The above matrices obey the commutation relation
\begin{equation}
[a_{q},\hat{n}]=a_{q},~~~[a_{q}^{\dag},\hat{n}]=-a_{q}^{\dag}
\end{equation}
but the commutation relations of the operators $~a_{q}$ and
$~a_{q}^{\dag}$ do not coincide with the boson commutation relations.
Eq. (72) is replaced by
\begin{equation}
[a_{q},a_{q}^{\dag}] = F(\hat{n})
\end{equation}
where the function $F(\hat{n})$ has the form
\begin{equation}
F(\hat{n}) = \frac{\sinh \lambda (\hat{n}+1)
-\sinh \lambda \hat{n}}{\sinh \lambda }.
\end{equation}
For $~\lambda =0$ (84) reduces to (72). In addition
to the above commutation relation there exists the reordering relation
\begin{equation}
a_{q}a_{q}^{\dag}-qa_{q}^{\dag}a_{q}=q^{-\hat{n}}
\end{equation}
which usually is taken as the definition of q--oscillators.

It is worthy noting that the operators $~a_{q}$ and $~a_{q}^{\dag}$
can be expressed in terms of the operators $a$ and $a^{\dag}$ (see,
for example, \cite{sol1})
\begin{equation}
a_{q}= af(\hat{n}),~~~a_{q}^{\dag}=f(\hat{n})a^{\dag}
\end{equation}
where
\begin{equation}
f(\hat{n})=\sqrt{\frac{\hat{n}_q}{\hat{n}}}.
\end{equation}
The comparison of formulae (45) and (87) shows the complete analogy of
the quantum q--oscillator and the classical q--oscillator discussed in
previous sections. We have also
\begin{equation}
\hat{n}_{q} = a_{q}^{\dag}a_{q}
\end{equation}
and
\begin{equation}
[a_{q},\hat{n}_{q}] = F(\hat{n}) a_{q},
{}~~~~~~~~~~~~~~[ a_{q}^{\dag}, \hat{n}_q] = - a_{q}^{\dag} F(\hat{n}).
\end{equation}

In the Schr\"{o}dinger representation the evolution operator of the
harmonic oscillator
\begin{equation}
U(t)=\exp \left[ -i\omega\, \frac{(a^{\dag}a + aa^{\dag})t}{2} \right]
\end{equation}
gives the possibility to find out explicitly linear integrals of motion which
depend on time
\begin{eqnarray}
A(t)&=&U(t)aU^{-1}(t)=e^{i \omega t}a,\nonumber\\
A^{\dag}(t)&=&U(t)a^{\dag}U^{-1}(t)=e^{-i \omega t}a^{\dag}.
\end{eqnarray}
The matrices of the integrals of motion (92) in Fock basis may be
obtained from the equations
\begin{eqnarray}
A(t)|n\rangle &=&e^{i \omega t} \sqrt{n} |n-1\rangle,\nonumber\\
A^{\dag}(t) |n\rangle &=&e^{-i \omega t} \sqrt {n+1} |n+1\rangle .
\end{eqnarray}
Let us now introduce the Hamiltonian
\begin{equation}
\hat{H}=\omega \frac{a_{q}a_{q}^{\dag} + a_{q}^{\dag}a_{q}}{2}
\end{equation}
for which the evolution operator takes the form
\begin{equation}
U_{q}(t)=\exp \left[ -i\omega t\frac{(a_{q} a_{q}^{\dag} +
a_{q}^{\dag}a_{q})}{2} \right].
\end{equation}
We have for the integrals of motion
\begin{equation}
A_{q}(t)=U_{q}(t)a_{q}U_{q}^{-1}(t),
{}~~~~~A_{q}^{\dag}(t)=U_{q}(t)a_{q}^{\dag}U_{q}^{-1}(t)
\end{equation}
the following explicit matrix expressions \cite{sol1}
\begin{eqnarray}
A_{q}(t)&=&\left(\begin{array}{cccc}
0 & ~~\sqrt{1_{q}}e^{i(1_{q}-0_{q})\omega t}~~ & 0 &  \ldots \\
0 & 0 & ~~\sqrt{2_{q}}e^{i(2_{q}-1_{q})\omega t}~~ & \ldots \\
{}~~\ldots~~ & \ldots & \ldots & \ldots\\
\end{array}\right),\nonumber\\
A_{q}^{\dag}(t)&=&\left(\begin{array}{cccc}
0 & 0 & 0 & \ldots\\
\sqrt{1_{q}}e^{-i(1_{q}-0_{q})\omega t}~~  & 0 & 0 & \ldots\\
0 & \sqrt{2_{q}}e^{-i(2_{q}-1_{q})\omega t}~~& 0 & \ldots\\
{}~~\ldots~~ & \dots & \dots & \ldots\\
\end{array}\right).
\end{eqnarray}
These operators are the generalizations of the linear integrals of motion
(92) to the case of nonlinear Hamiltonian. This result is a generalization
for q--oscillators of such integrals of motion for usual and parametric
quantum oscillator which have been found in \cite{mal70} and discussed for
constructing coherent states in \cite{malman79} and \cite{dod83}.
{}~~\\
{}~~\\
{}~~\\
{\bf ANALOGY OF CLASSICAL AND QUANTUM DEFORMATIONS}
{}~~\\

By using the example of the oscillator we clarify now the analogy which
exists in the suggested approach to deformation of classical systems and
quantum ones. Now we recall what has been done in the previous Section
for q--deformed quantum oscillator to make clear the connection with
the procedure of deforming the classical oscillator of first sections.

Equations of motion for the harmonic oscillator amplitude $~a$ we rewrite
for another operator
$$A=h(a\dag a)a$$
where $~h$ is a real function, and its hermite conjugate $~A\dag
=~a\dag ~h(a\dag a)$ in the same form
$$\dot A=-i\omega A,~~~~~\dot A\dag =i\omega A\dag .$$
We have in our Hilbert space the vacuum state $~\Psi _{0}$ which satisfies
$$a\Psi _{0}=0,~~~~~~A\Psi_{0}=0.$$
Thus we could construct two bases in the vector space. One is the standard
basis
$$~\Psi _{n}=\frac {a^{\dag n}}{\sqrt {n!}}\Psi _{0}$$
which is orthonormal in standard scalar product
$$<\Psi_{n}|\Psi _{m}>=\delta _{nm}.$$
Another basis is constructed using the operator $~A\dag$
$$\widetilde \Psi _{n}=\frac {(A\dag )^{n}}{\sqrt {n!}}\Psi_{0}.$$
We define new scalar product in the same vector space which is given by
$$<\widetilde \Psi _{n}|\widetilde \Psi _{m}>=\delta _{nm}.$$
The adjoint with respect to this new scalar product need not coincide
with the old one. We can define the operators
$$b^{*}\widetilde \Psi _{n}=\sqrt {n+1}\widetilde \Psi _{n+1},~~~~~
b\widetilde \Psi _{n}=\sqrt {n}\widetilde \Psi _{n-1}$$
where $~*$ means the adjoint in the new scalar product. These operators
satisfy the commutation relations $~[b,~b^{*}]=1$. Taking the Hamiltonian
$~H=\omega b^{*}b$ we have for the operators $~b,~b^{*}$ the equation of
motion of the harmonic oscillator. Thus for one and
the same vector space we have possibility to introduce two Hilbert space
structures. As for the dynamics we have, like in the classical case, two
different descriptions. Very much as we did for the classical case we can
use the new Hamiltonian and the old commutator relations to get a "deformed"
dynamics. As for the partition function, similarly to the classical case,
we can use the trace defined via the two different scalar products to get
the same result, i. e. the partition function depends only on the dynamics
and not on the particular Hamiltonian description we use.
Either we change the Hamiltonian and for the old scalar product we
obtain new dynamics. Or changing Hamiltonian and simultaneously the scalar
product we obtain the same dynamics. For partition function in such case we
have the same value that was for nondeformed oscillator.
{}~~\\
{}~~\\
{}~~\\
{\bf DEFORMED PLANCK DISTRIBUTION}
 ~~\\

In this Section we will discuss what physical consequences may be found
if the considered q--nonlinearity influences the vibrations of the real
field mode oscillators like, for example, electormagnetic fields ones
or the oscillations of the nuclei in polyatomic molecules. First of all
this nonlinearity changes the specific heat behaviour. To show this we
have to find the partition function for a single q--oscillator
corresponding to the Hamiltonian $~H=\hat n_{q}$
\begin{equation}
Z(T)=\sum_{n=0}^{\infty} \exp(-\beta n_{q})
\end{equation}
where the variable$~\beta $
is the function of the temperature $~T^{-1}$. The evaluation of the quantum
partition function of the q--oscillator yields for the specific heat that it
decreases for $~T\rightarrow \infty $ as
\begin{equation}
C\propto \frac{1}{\ln~T}.
\end{equation}
 Thus the behaviour of the specific heat of the q--oscillator found in
\cite{sol1} for $~\lambda \ll 1$ is different from the behaviour
of the usual oscillator in the high temperature limit. This property may
serve for an experimental check of the existence of vibrational
nonlinearity of the q-oscillator fields.

q--Deformed Bose distribution can be obtained by the same
method starting from the Hamiltonian
$~H=~\frac {1}{2}\{a^{\dag}_q,a_q\}_{+}$ and one obtains \cite{sol1}
\begin{equation}
<n>=\bar n_{0}-\beta \frac{\lambda ^{2}}{6}\left[
\frac{1}{2}((\bar {n^2})_0-(\bar n)^{2}_0)+\frac{3}{2}((\bar {n^3})_0
-\bar {n}_{0}(\bar {n^2})_0)
+(\bar {n^4})_0-\bar {n}_{0}(\bar {n^3})_0)\right]
\end{equation}
in which $~\bar{n}_0$ is the usual Bose distribution function and
\begin{equation}
(\bar {n^k})_0 = 2\sinh \frac{\beta}{2} \sum_{n=0}^{\infty}
 n^{k} e^{-\beta (n+1/2)}.
\end{equation}

Calculating the partition function for small q--nonlinearity parameter
we have also the following q--deformed Planck distribution formula
\begin{equation}
<n>=\frac {1}{e^{\hbar \omega /kT}-1}-\lambda ^{2}\frac {\hbar \omega}
{kT}\frac {e^{3\hbar \omega /kT}+4e^{2\hbar \omega /kT}
+e^{\hbar \omega /kT}}{(e^{\hbar \omega /kT}-1)^{4}}.
\end{equation}
It means that q--nonlinearity deforms the black body radiation formula
\cite{sol1}.

One can write down the high and low temperature approximations for
the deformed Planck distribution formula \cite{sol3}.
For small temperature the behaviour of the deformed Planck distribution
differs from the usual one
\begin{equation}
<n>-\bar n_{0} = -\lambda ^{2}\frac {\hbar \omega }{kT}
e^{-\hbar \omega /kT}.
\end{equation}
For the high temperature the nonlinear correction to the usual Planck
distribution also depends on temperature
\begin{equation}
<n>-\bar n_{0} = - 6\lambda ^{2}
(\frac {\hbar \omega }{kT})^{-3}.
\end{equation}
As it was seen, the discussed q-nonlinearity produces a correction to
Planck distribution formula and also this may be subjected to an
experimental test.

As it was suggested in \cite{sol2} the q--nonlinearity of the field
vibrations produces blue shift effect which is the effect of the
frequency increase with the field intensity. For small nonlinearity
parameter $~\lambda $ and for large quantity of photons $~n$ in
a given mode the relative shift of the light frequency is
$$\frac {\delta \omega }{\omega }
=\frac {\lambda ^{2}}{2}(n-\frac {1}{3}).$$
This phenomenon of possible existence of the q--nonlinearity may
be essential for the models of the early stage of the Universe.

Another possible phenomenon related to the q--nonlinearity was
considered in \cite{marman} where it was shown that if one deforms
the electrostatics equation using the method of deformed creation
and annihilation operators the formfactor of a point charge appears
due to q--nonlinearity.
{}~~\\
{}~~\\
{}~~\\
{\bf NONLINEAR KLEIN--GORDON EQUATION}
{}~~\\

To demonstrate how the q--nonlinearity may appear in Klein--Gordon
equation we start from the
consideration of usual Klein--Gordon equation with mass equal to zero
($~c=1$)
\begin{equation}
(\frac{\partial ^2}{\partial t^2}-\Delta)\varphi (\hbox {\bf x},t)=0.
\end{equation}
Let us take the plane wave solutions of the equation, i.e., we represent
the field $~\varphi (\hbox {\bf x},t)$ in the form
\begin{equation}
\varphi (\hbox {\bf x},t)
=\int \varphi (\hbox {\bf k},t)\exp (i \hbox {\bf kx})~d\hbox {\bf k},
\end{equation}
where Fourier amplitude
\begin{equation}
\varphi (\hbox {\bf k},t)=\frac{1}{(2\pi )^3} \int \varphi (\hbox {\bf x},t)
\exp (-i\hbox {\bf kx})~d\hbox {\bf x},
\end{equation}
plays the role of new coordinate. It satisfies the integral equation
\begin{equation}
\int \ddot \varphi (\hbox {\bf k},t)\exp (i\hbox {\bf kx})~d\hbox {\bf k}
=\int (-k^2)\varphi (\hbox {\bf k},t)\exp (i\hbox {\bf kx})~d\hbox {\bf k}.
\end{equation}
This integral equation is equivalent to the differential one
\begin{equation}
\ddot \varphi (\hbox {\bf k},t)+k^2\varphi (\hbox {\bf k},t)=0,
\end{equation}
which is an infinite system of decoupled oscillators with frequencies
$~\omega ^2=k^2$. According to suggested procedure we replace this equation
by
\begin{equation}
\ddot \varphi (\hbox {\bf k},t)
+k^{2}f_{q}^{2}(\mu )\varphi (\hbox {\bf k},t)=0.
\end{equation}
Here the new frequency of $~\hbox {\bf k}$--th mode appeared
\begin{equation}
\omega ^{2}=k^{2}f_{q}^{2}(\mu),
\end{equation}
where the choice of parameter $~\mu$ determines the deformation. According
to our ideology it is possible to take it to be an integral of motion of
the Klein--Gordon equation. There exists a common integral of motion which
is the full number of the scalar field quanta, and we take it to be equal
$~\mu$. As well as for the wave equation case the parameter $~\mu$ behaves
as constant for any choice of the initial conditions
$~\varphi (\hbox {\bf x},t=0),~~\dot \varphi (\hbox {\bf x},t=0)$.
Due to this the solution to the nonlinear
mode--vibration equation is the following
\begin{eqnarray}
\varphi (\hbox {\bf k},t)&=&\frac{1}{2}\varphi (\hbox {\bf k},0)
\{\exp [i|k|f_{q}(\mu )]t+
\exp [-i|k|f_{q}(\mu )]t\}\nonumber\\
&+&\frac{1}{2i}\dot \varphi (\hbox {\bf k},0)\{\exp [i|k|f_{q}(\mu )]t
-\exp [-i|k|f_{q}(\mu )]t\}\frac {1}{i|k|f_{q}(\mu )},
\end{eqnarray}
Here
\begin{eqnarray}
\varphi (\hbox {\bf k},0)&=&\frac{1}{(2\pi )^3}\int_{-\infty}^{\infty}
\varphi (\hbox {\bf x},0)\exp (-i\hbox {\bf kx})~d\hbox {\bf x},\nonumber\\
\dot \varphi (\hbox {\bf k},0)&=&\frac{1}{(2\pi )^3}\int_{-\infty}^{\infty}
\dot \varphi (\hbox {\bf x},0)\exp (-i\hbox {\bf kx})~d\hbox {\bf x}.
\end{eqnarray}
Thus the given initial conditions $~\varphi (\hbox {\bf x},0),
{}~~\dot \varphi (\hbox {\bf x},0)$ determine also
$~\varphi (\hbox {\bf k},0)$, $~\dot \varphi (\hbox {\bf k},0)$,
and $~\mu $. In terms of these values we have the
$~\hbox {\bf k}$--th
mode solution $~\varphi (\hbox {\bf k},t)$ and the solution to nonlinear
q--deformed  Klein--Gordon equation
\begin{equation}
(\frac{\partial ^2}{\partial t^2}-f_{q}^{2}(\mu)\Delta)
\varphi (\hbox {\bf x},t)=0.
\end{equation}
The constant of motion $~f_{q}(\mu)$ plays the role of signal velocity.
{}~~\\
{}~~\\
{}~~\\
{\bf q--DEFORMED ELECTRODYNAMICS}
{}~~\\

We will consider the system of usual linear Maxwell equations in vacuum
for the field
$~\hbox {\bf E}(\hbox {\bf x},t),~\hbox {\bf H}(\hbox {\bf x},t)$
expressed in terms of the potentials
$~\hbox {\bf A}(\hbox {\bf x},t),~\varphi (\hbox {\bf x},t)$
\begin{eqnarray}
\hbox {\bf E}&=&-\frac {1}{c}\frac {\partial \hbox {\bf A}}{\partial t}
-\frac {\partial \varphi }{\partial \hbox {\bf x}},\nonumber\\
\hbox {\bf H}&=&rot~\hbox {\bf A},
\end{eqnarray}
in the form ($~c=1$)
\begin{equation}
(\frac {\partial ^{2}}{\partial t^{2}}-\Delta )\varphi (\hbox {\bf x},t)=0,
\end{equation}
\begin{eqnarray}
(\frac {\partial ^{2}}{\partial t^{2}}
-\Delta )\hbox {\bf A}(\hbox {\bf x},t)
&=&0,\nonumber\\
\frac {\partial \varphi }{\partial t}+div~\hbox {\bf A}&=&0.
\end{eqnarray}
The suggested approach for scalar field of previous Section may be
applied if one reexpresses the fields
$~\varphi (\hbox {\bf x},t),~\hbox {\bf A}(\hbox {\bf x},t)$
in Fourier basis
\begin{eqnarray}
\varphi (\hbox {\bf x},t)&=&\int \varphi (\hbox {\bf k},t)
\exp (i\hbox {\bf kx})~d\hbox {\bf k},\nonumber\\
\hbox {\bf A}(\hbox {\bf x},t)&
=&\int \hbox {\bf A}(\hbox {\bf k},t)\exp (i\hbox {\bf kx})~d\hbox {\bf k}.
\end{eqnarray}
Then we have
\begin{eqnarray}
\ddot \varphi (\hbox {\bf k},t)+k^{2}\varphi (\hbox {\bf k},t)&=&0,\nonumber\\
\ddot {\hbox {\bf A}}(\hbox {\bf k},t)+k^{2}\hbox {\bf A}(\hbox {\bf k},t)&=&0
\end{eqnarray}
together with gauge constraint
\begin{equation}
\dot \varphi (\hbox {\bf k},t)+i\hbox {\bf kA}(\hbox {\bf k},t)=0.
\end{equation}
The approach is based on the taking into account that the equations
describing the linear forced oscillators of the electromagnetic field
are deformed due to dependence of the frequency of the oscillations
on the energy of the electromagnetic vibrations. As in the case of the
scalar field we will introduce the total "number of quanta" of the
defiormed electromagnetic field $~\mu$
which is integral of motion. Then in this case we have
the deformed nonlinear equations of vibrations
\begin{eqnarray}
\ddot \varphi (\hbox {\bf k},t)+f_{q}^{2}(\mu )k^{2}\varphi (\hbox {\bf k},t)
&=&0,\nonumber\\
\ddot {\hbox {\bf A}}(\hbox {\bf k},t)+f_{q}^{2}(\mu )k^{2}
\hbox {\bf A}(\hbox {\bf k},t)&=&0,
\end{eqnarray}
The sense of the function $~f_{q}(\mu)$ is the signal velocity.Due to this we
have to reparametrize the gauge condition to become
\begin{equation}
f_{q}^{-1}(\mu )\varphi (\hbox {\bf k},t)
+i\hbox {\bf kA}(\hbox {\bf k},t)=0.
\end{equation}
After this we could reconstruct Maxwell equations in space--time analogously
to the wave equation case. It should be noted that parameter $~\mu $ in the
Klein--Gordon and Maxwell equations is given by Eq. (64). The limit of
electrostatics in the suggested type of q--deformation coincides with the
usual electrostatics described by the Laplace equation (compare with
\cite{marman}). It would be  interesting to take into account interaction
of the deformed electromagnetic field with the sources, but it will be taken
up elsewhere.

\end{document}